\documentclass[hidelinks,10pt]{article}
\usepackage{amsmath,amssymb,latexsym,multirow,color,titlesec}
\usepackage[nosort]{cite}
\usepackage{mathtools}
\usepackage{graphicx, fancyhdr}    
\usepackage{amsfonts}              
\usepackage{subcaption}
\usepackage{amsthm,amssymb}        
\usepackage[font=small,labelfont=bf]{caption}
\usepackage{color, booktabs, amscd, hyperref}
\usepackage{float,graphicx,ulem}
\usepackage{rotating,longtable}
\usepackage{pdflscape,tikz} 
\floatstyle{plaintop}
\restylefloat{table}
\newcommand{\ds}{\displaystyle}
\titleformat{\paragraph}
{\normalfont\normalsize\bfseries}{\theparagraph}{1em}{}
\titlespacing*{\paragraph}
{0pt}{3.25ex plus 1ex minus .2ex}{1.5ex plus .2ex}

\begin{document}

\title{Atom-specific persistent homology and its application to protein flexibility analysis }
\author{David Bramer$^{1}$ and
Guo-Wei Wei$^{1,2,3}$ \footnote{ Address correspondences  to Guo-Wei Wei. E-mail:wei@math.msu.edu}\\
$^1$ Department of Mathematics \\
Michigan State University, MI 48824, USA\\
$^2$  Department of Biochemistry and Molecular Biology\\
Michigan State University, MI 48824, USA \\
$^3$ Department of Electrical and Computer Engineering \\
Michigan State University, MI 48824, USA \\
}

\maketitle
\newcommand{\rd}[1]{{\textcolor{red}{#1}}}
\begin{abstract}
Recently, persistent homology has had tremendous success in biomolecular data analysis. It works by examining the topological relationship or connectivity of a group of atoms in a molecule at a variety of scales,  then rendering a family of topological representations of the molecule. However, persistent homology is rarely  {employed} for the analysis of atomic properties, such as biomolecular flexibility analysis or B factor prediction. This work introduces atom-specific persistent homology to provide a local atomic level representation of a molecule via a global topological tool. This is achieved through the construction of a pair of conjugated sets of atoms and corresponding conjugated simplicial complexes, as well as conjugated topological spaces. The difference between the topological invariants of the pair of conjugated sets is measured by Bottleneck and Wasserstein metrics and leads to an atom-specific topological representation of individual atomic properties in a molecule.  Atom-specific topological features are integrated with various machine learning algorithms, including gradient boosting trees and convolutional neural network for protein thermal fluctuation analysis and B factor prediction. Extensive numerical results indicate the proposed method provides a powerful topological tool for analyzing and predicting localized information.

\end{abstract}
\newpage

\tableofcontents
\newpage

\section{Introduction}

In recent years tools from topology have been successfully applied to protein analysis \cite{KLXia:2014c, Gameiro:2014,KLXia:2015b, ZXCang:2015,Kovacev-Nikolic:2016,xia2018persistent}. Topology  offers one of highest level of abstractions of geometric data and allows one to infer high dimensional structure from low dimensional representations. However, conventional topology oversimplifies geometry and thus lacks descriptive power for most real world problems. Persistent homology (PH) overcomes this difficulty by introducing a filtration parameter that describes the geometry in terms of a family of  Betti numbers at various scales known as a barcode  \cite{Frosini:1999, Edelsbrunner:2002,Zomorodian:2005,Zomorodian:2008}.
Indeed, three dimensional (3D) protein spatial information from a protein data bank (PDB) file can be converted into a family of simplicial complexes. One can apply tools from algebraic topology to  {convert} structural information into global topological invariants that provide a useful representation of biomolecular properties \cite{yao2009topological}. However, for quantitative biomolecular analysis and prediction, persistent homology  {alone} neglects chemical and biology information. Element-specific persistent homology has been introduced to incorporate chemical and biological information into topological invariants \cite{ZXCang:2017a,ZXCang:2017b}. Similarity and differences between barcodes from different molecules can be measured by Wasserstein \cite{cohen2010lipschitz} and/or Bottleneck \cite{CEH07} distances. However, the previous applications of persistent homology and element-specific persistent homology are for the modeling and prediction of  molecule-level   thermodynamical or structural properties, such as protein-ligand binding affinities  \cite{ZXCang:2017b},   protein folding free energy changes  upon mutations \cite{ZXCang:2017a,ZXCang:2017c}, drug toxicity \cite{KDWu:2018a},  solubility, partition coefficient \cite{KDWu:2018b}, and drug virtual screening (ligand and decoy classification) \cite{ZXCang:2018a}. Essentially, topology is a global tool that examines the connectivity and relationship among many atoms in a neighborhood as a whole. High dimensional topological invariants, such as Betti 1 and Betti 2, describe the collective behavior of many atoms.  Therefore, it is not clear how to represent atomic level property, such as the B factor of an atom, by persistent homology.  
 
 In proteins, beta factor (B factor) or  (Debye-Waller factor is a measure of the attenuation of  X-ray scattering caused by thermal motion. The strength of the thermal motion of an atom is theoretically proportional to its B factor during the structure determination from X-ray  diffraction data. It is well known that biomolecular flexibility provides an important link between its structure and function. In particular, it has been shown that intrinsic structural flexibility correlates to meaningful protein conformational variations, reactivity and enzymatic function \cite{JMa:2005}. As such, the accurate prediction of protein B-factor is essential to our understanding of protein structure, function and dynamics  \cite{Frauenfelder:1991}.


Early methods used to predict protein B factor were derived from Hooke's Law and are known as elastic mass-and-spring networks. In these models, alpha carbons (C$_\alpha$) of biological macromolecules are treated as a mass and spring network and motions are predicted based on a harmonic potential. Given a protein, each C$_\alpha$ is represented as a node in the network and edges are weighted based on a potential function. Nodes are connected by an edge if they fall within a pre-defined euclidean cutoff distance. This captures the local covalent and non-covalent interactions between an individual atom and nearby atoms. One of the first mass-and-spring methods used for protein B factor prediction is normal mode analysis (NMA). Like most B factor prediction methods, NMA is independent of time and uses a Hamiltonian interaction matrix. Eigenvalues of the matrix system correspond to characteristic frequencies of the protein and these frequencies correlate with protein B factors. Low-frequency modes correlate with cooperative motion and can be useful for hinge detection and domain motion. NMA has also been successfully implemented to understand the deformation of supramolecular complexes. \cite{JMa:2005,Tasumi:1982,Brooks:1983,Levitt:1985}

Elastic network model (ENM) was introduced as a more efficient model that significantly reduces computational cost compared to NMA through the use of a simplified spring network \cite{Tirion:1996}. A specific example is  anisotropic network model (ANM) \cite{Atilgan:2001}. Gaussian network model (GNM) further reduces the computational cost by ignoring the anisotropic motion, rendering a more accurate method for protein C$_{\alpha}$ B factor analysis \cite{Bahar:1997,Bahar:1998,Haliloglu:1997}.

All of the aforementioned methods depend on matrix diagonalization, which has the computational complexity of $\mathcal{O}(N^3)$, where is the number of matrix atoms involved in the analysis.  Recently, flexibility and Rigidity Index (FRI) methods have  {been} proposed as a geometric graph approach to further reduce the computational cost. FRI methods rely on constructing a distance matrix using radial basis functions to scale atom to atom distance non-linearly \cite{KLXia:2013f}. All versions of FRI produce a flexibility index, that correlates to the B factor, for each C$_\alpha$. Several versions of FRI have been developed. Among them, fast FRI (fFRI) is of $\mathcal{O}(N)$ in computational complexity \cite{Opron:2014}. FRI methods are also more accurate than all of the earlier  algebraic graph-based methods. Additionally,  anisotropic FRI (aFRI) provides   high quality anisotropic motion analysis  \cite{Opron:2014}. Moreover, using several radial basis functions with different parametrizations,   the multiscale flexibility rigidity index (mFRI) can successfully capture multiscale atomic interactions \cite{Opron:2015a}.

More recently, the authors introduced a multiscale weighted colored graph (MWCG) model. The MWCG is another geometric graph theory model that has been shown to be the best B factor prediction model to date. First, element-specific interaction subgraphs are constructed based on selected atomic interactions between certain element types. Atoms are represented as graph nodes and subgraphs are generated using pairs of atoms of certain elements (e. g., carbon, nitrogen, oxygen, sulfur). A centrality metric that uses radial basis functions is applied to pairwise interactions in each subgraph. By varying the parametrization of the radial basis functions the MWCG model can capture multiple protein interaction scales. MWCG is unique in its ability to utilize both element specific and multiscale interactions for improved B factor prediction \cite{DBramer:2018a}. Most recently, MWCG is incorporated with machine learning algorithms for across-protein blind predictions of protein B factors \cite{DBramer:2018b}.

The objective of the present work is to extend the utility of persistent homology for atomic level property modeling and prediction. To this end, we introduce atom-specific persistent homology (ASPH) to create a local atomic representation of an atom using a global topological tool in a novel way. Specifically, ASPH  constructs  a pair of conjugated sets of point clouds or atoms centered around the atom of interest. The first set of a pair of conjugated sets of atoms for a given atom is selected by a local sphere of radius $r_c$ around the atom of interest. The second set of atoms is defined by excluding the atom of interest in the first set.  Conjugated simplicial complexes, conjugated chain groups, conjugated homology groups as well as conjugated persistence barcodes or diagrams are induced by an identical filtration. Conjugated persistence barcodes are compared with Bottleneck and Wasserstein metrics. The resulting distance provides a global topological representation of the localized atomic property, such as protein flexibility analysis and atomic-level protein B-factor information.   
Obviously, the proposed atom-specific topology can be applied to a wide variety of chemical and biological problems where atomic properties are measured, such as the chemical shifts of nuclear magnetic resonance (NMR), the B-factors of X-ray structure determination, and the shift and line broadening of other atomic spectroscopy.  
 
We focus on protein C$_\alpha$ B-factor prediction but the approach provided in this work is a general framework that can be used to predict B factors of any atom in a protein. First, we use the generated atom-specific persistent homology features to fit B factors within a given protein using   linear least squares  {minimization}. Then the atom-specific persistent homology features are combined with other local and global protein features to construct machine learning models for the blind prediction of protein B factors across different proteins. Additionally,  image-like multiscale atom-specific persistent homology features are generated using an early technique \cite{KLXia:2015c}.  These image like features, together with other features, are fed into convolutional neural networks (CNN).  Training and validation are carried out using a large and diverse set of proteins from the protein data bank (PDB). The proposed method offers some of the best results for blind B factor predictions of a set of 364 proteins.


\section{Methods and algorithms}\label{sect:Theory}

\subsection{Atom-specific persistent homology}
\subsubsection{Overview}
Topology describes (continuous) objects in terms of topological invariants, i.e., Betti numbers.  Betti-0, Betti-1, and Betti-2  which  can be interpreted as connected components, rings,  cavities, etc. Table \ref{tab:betti} provides examples of the  Betti numbers of a point, circle, sphere, and torus.

\newcommand{\wid}{0.75\textwidth}
\begin{table}
\caption{Topological invariants displayed as Betti numbers. Betti-0 represents the number of connected components,  Betti-1 the number of tunnels or circles, and Betti-2 the number of cavities or voids. Two auxiliary rings are added to the torus to illustrate that its Betti-1=2.}
\label{tab:betti}
\begin{tabular}{|c|c|c|c|c|}\hline
\quad &\begin{minipage}{0.2\textwidth}\centering\vspace{1mm}
\includegraphics[width=\wid]{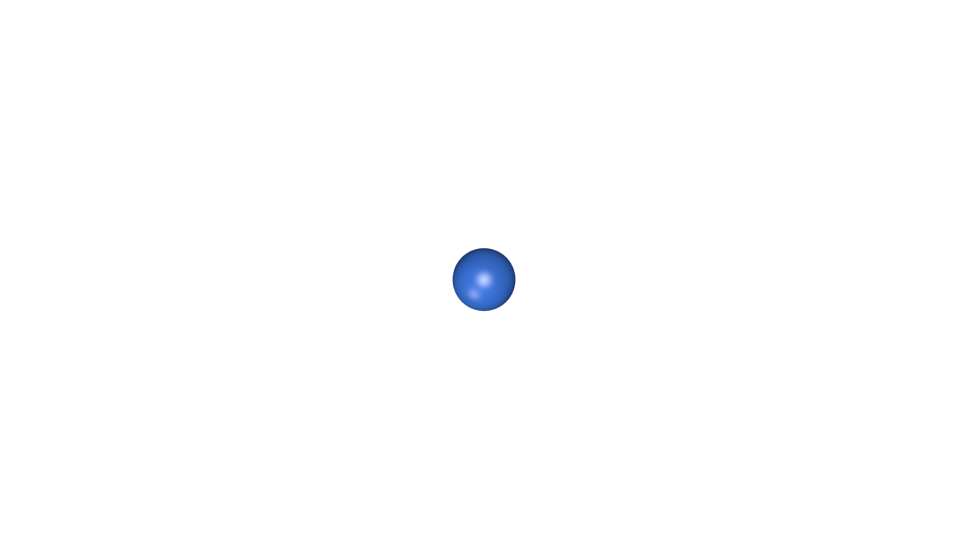}
\end{minipage}&
\begin{minipage}{0.2\textwidth}\centering\vspace{1mm}
\includegraphics[width=\wid]{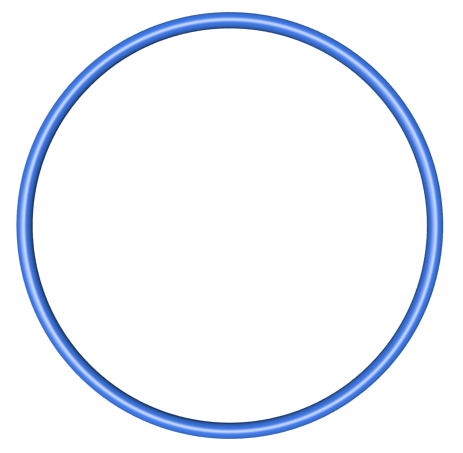}
\end{minipage}&
\begin{minipage}{0.2\textwidth}\centering\vspace{1mm}
\includegraphics[width=\wid]{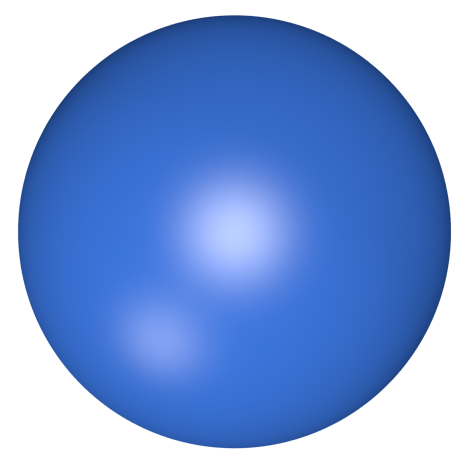}
\end{minipage}&
\begin{minipage}{0.2\textwidth}\centering\vspace{1mm}
\includegraphics[width=1\textwidth]{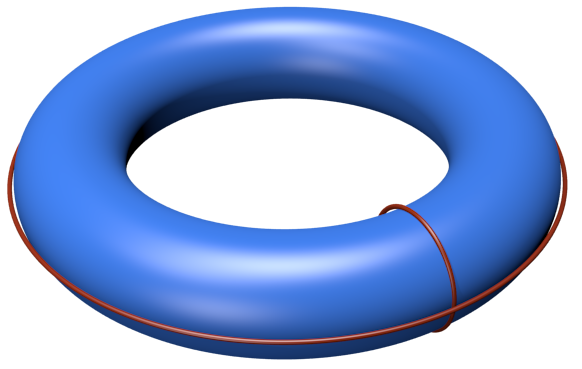}
\end{minipage}\\ \hline
Example&Point & Circle & Sphere & Torus \\ \hline
Betti-0 & 1&1&1&1\\ \hline
Betti-1 & 0&1&0&2\\ \hline
Betti-2 & 0&0&1&1\\ \hline
\end{tabular}
\end{table}

\begin{figure}[H]
\centering
\begin{minipage}{0.24\textwidth}
\begin{tikzpicture}
\hspace{0.5cm}\draw[black,thick,fill=red] (0,3) circle (0.1cm);
\end{tikzpicture}
\end{minipage}
\begin{minipage}{0.24\textwidth}
\begin{tikzpicture}
\draw[black, ultra thick,black] (1,2.5) -- (0,0);
\draw[black,thick,fill=red] (0,0) circle (0.1cm);
\draw[black,thick,fill=red] (1,2.5) circle (0.1cm);
\end{tikzpicture}
\end{minipage}
\begin{minipage}{0.24\textwidth}
\begin{tikzpicture}
\draw[black,ultra thick,fill=lightgray] (0,0) -- (1.5,0) -- (2,2.5) -- cycle;
\draw[black, thick,fill=red] (0,0) circle (0.1cm);
\draw[black,thick,fill=red] (1.5,0) circle (0.1cm);
\draw[black,thick,fill=red] (2,2.5) circle (0.1cm);
\end{tikzpicture}
\end{minipage}
\begin{minipage}{0.24\textwidth}
\begin{tikzpicture}
\draw[black,ultra thick,fill=lightgray] (0,0) -- (1,2) -- (1.25,-1) -- cycle;
\draw[black,ultra thick,fill=gray] (1,2) -- (1.25,-1) -- (2.5,0.1) -- cycle;
\draw[black,ultra thick,dashed] (0,0) -- (2.5,0.1);
\draw[black,thick,fill=red] (1.25,-1) circle (0.1cm);
\draw[black,thick,fill=red] (0,0) circle (0.1cm);
\draw[black,thick,fill=red] (1,2) circle (0.1cm);
\draw[black,thick,fill=red] (2.5,0.1) circle (0.1cm);
\end{tikzpicture}
\end{minipage}
\caption{From left to right an example of a 0-simplex, 1-simplex, 2-simplex, and 3-simplex.}
\label{fig:simplex}
\end{figure}
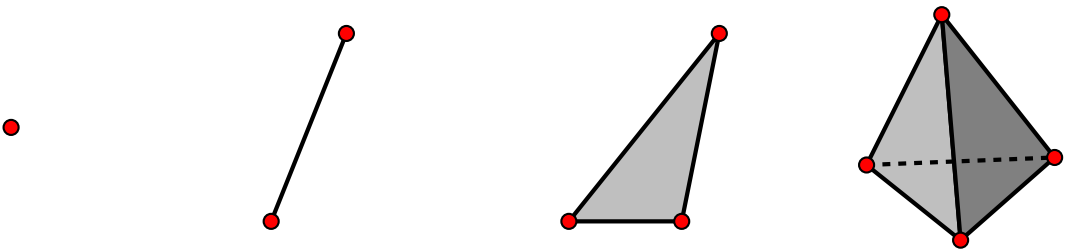

 Given discrete data points, such as a point cloud or the set of atoms in a molecule, we use simplicial complexes to describe the topological relationship, or connectivity of the point cloud, to systematically identify topological invariants.  First, a few simplicial complexes, as shown in Figure \ref{fig:simplex}, are made up of vertices, edges, triangles, and tetrahedrons, denoted 0-simplex, 1-simplex, 2-simplex, and 3-simplex, respectively. Homology groups constructed from simplicial complexes give rise topological invariants.   {Given discrete dataset or a set of protein atoms, nontrivial topological information is generated by persistent homology. This introduces a filtration parameter to create a family of simplexes, which leads to a family of simplicial complexes, homology groups and associated topological invariants.} By continuously varying the filtration parameter over an interval, the topological relationship among a given set of atoms is systematically reset, rendering a family of homology groups and corresponding topological invariants, which can be plotted as a persistence diagram, or a set of  barcodes. Both  persistence diagrams and barcodes  record the birth and death (appearance and cessation) of Betti numbers during the filtration process. Many simplicial complex definitions, which determine the rules of  the corresponding  topological relationship, have been proposed. Commonly used  definitions  include Vietoris-Rips (VR) complex, $\check{\mbox{C}}$ech complex, and alpha complex. 

Persistent homology allows the extraction of topological invariants that are embedded in the high dimensional data space of biomolecules.  The resulting topological invariants over the filtration, i.e., persistence diagrams or persistence barcodes of different molecules can be compared using Bottleneck and Wasserstein distances. 

The goal of atom-specific persistent homology is to extract topological information of a given atom in a molecule. To embed local atomic information into a global topological description, we construct a pair of conjugated sets of point clouds, namely the original dataset and a datset excluding the atom of interest. The Bottleneck and Wasserstein distances between these two persistence diagrams reveal the desirable topological information of the given atom.

\subsubsection{Simplex and  simplicial complex}
A (geometric) simplex is a generalization of a triangle or tetrahedron to arbitrary dimensions. A $k$-simplex is a convex hull of $k+1$ vertices represented by a set of affinely independent points \begin{equation}
\sigma=\{\lambda_0u_0+\lambda_1u_1+\ldots+\lambda_ku_k\mid \sum\lambda_i=1,\lambda_i\geq0,i=0,1,\ldots,k\},
\end{equation}
where $\{u_0,u_1,\ldots,u_k\}\subset\mathbb{R}^d$ with $d\geq k$ is the set of points, $\sigma$ is the $k$-simplex, and constraints on $\lambda_i$'s ensure the formation of a convex hull. An affinely independent combination of points can have at most $k+1$ points in $\mathbb{R}^k$. For example a 1-simplex is a line segment, a 2-simplex a triangle, and a 3-simplex a tetrahedron. A subset of the $k+1$ vertices of a $k$ simplex with $m+1$ vertices forms a convex hull in a lower dimension and is called an $m$-face of the $k$-simplex. An $m$-face is proper is $m<k$. The boundary of a $k$-simplex $\sigma$, is defined as the formal sum of its $(k+1)$ faces. Given as \begin{equation}
\partial_k\sigma=\displaystyle\sum_{i=0}^{k}
(-1)^i[u_0,\ldots,\hat{u}_i,\ldots,u_k],
\label{eq:1}
\end{equation} where $[u_0,\ldots,\hat{u}_i,\ldots,u_k]$ denotes the convex hull formed by vertices of $\sigma$ with the vertex $u_i$ being excluded and $\partial_k$ is called the boundary operator. A collection of finitely many simplicies forms a simplicial complex denoted by $\mathcal{K}$. All simplicial complexes satisfy the following conditions. \begin{enumerate}
\item Faces of any simplex in $\mathcal{K}$ are also simplices in $\mathcal{K}$.
\item The intersection of any two simplicies $\sigma_1,\sigma_2\in\mathcal{K}$ is a face of both $\sigma_1$ and $\sigma_2$.
\end{enumerate}

\subsubsection{Homology}
Given a simplicial complex $\mathcal{K}$, a $k$-chain $c_k$ of $\mathcal{K}$ is a formal sum of the $k$-simplices in $\mathcal{K}$ 
and is defined as $c_k=\sum a_i\sigma_i$ where $\sigma_i$ are the $k$-simplices and $a_i$'s coefficients. Generally, $a_i$ are element of a field such as $\mathbb{R}$, $\mathbb{Q}$, or $\mathbb{Z}_n$. Computationally, it is common to choose $a_i$ to be in $\mathbb{Z}_2$. The group of $k$-chains in $\mathcal{K}$, denoted $C_{k}$, forms an Abelian group under addition in modulo two. This allows us to extend the definition of the boundary operator introduced in Eq. (\ref{eq:1}) to chains.

The boundary operator applied to a $k$-chain $c_k$ is defined as \begin{equation}
 \partial_kc_k=\sum a_i\partial_k\sigma_i, \end{equation} where $\sigma_i$'s are $k$-simplices. The boundary operator is a map from $\mathcal{C}_k$ to ${\mathcal{C}}_{k-1}$, which is also known as a boundary map for chains. Note that in $\mathbb{Z}_2$, the boundary operator $\partial_k$ satisfies the property that $\partial_k\circ\partial_{k+1}\sigma = 0$ for any $(k + 1)$-simplex $\sigma$ following the fact that any $(k-1)$-face of $\sigma$ is contained in exactly two $k$-faces of $\sigma$. The chain complex is defined as a sequence of chains connected by boundary maps with decreasing dimension and is denoted
\begin{equation} \ldots \rightarrow \mathcal{C}_{n}(\mathcal{K} )\xrightarrow[]{\partial_{n}}\mathcal{C}_{n-1}(\mathcal{K})\xrightarrow[]{\partial_{n-1}} \ldots\xrightarrow[]{\partial_{1}}\mathcal{C}_{0} (\mathcal{K} )\xrightarrow[]{\partial_{0}}{}0.  
\end{equation} 
The $k$-cycle group and $k$-boundary group are then defined as kernel and image of $\partial_k$ and $\partial_{k+1}$ respectively, and \begin{eqnarray}
\mathcal{Z}_k&=& \mbox{Ker}\partial_k=\{c \in \mathcal{C}_k\mid\partial_kc = 0\},\\ \mathcal{B}_{k}&=& \mbox{Im}\partial_{k+1}=\{c\in \mathcal{C}_k | \exists d\in    \mathcal{C}_{k+1}: c= \partial_{k+1}d
\},\end{eqnarray}
where $\mathcal{Z}_k$ is the $k$-cycle group and $\mathcal{B}_k$ is the $k$-boundary group. Since $\partial_k\circ\partial_{k+1}= 0$, we have $\mathcal{B}_k\subset\mathcal{Z}_k\subset\mathcal{C}_k$. Then the $k$-homology group is defined to be the quotient group of the $k$-cycle group modulo the $k$-boundary group,
 \begin{equation}
\mathcal{H}_k=\mathcal{Z}_k/\mathcal{B}_k
\end{equation}where $\mathcal{H}_k$ is the $k$-homology group. The $k$th Betti number is defined to be rank of the $k$-homology group as $\beta_k={\rm rank}(\mathcal{H}_k)$.

\subsubsection{Filtration  and persistence}
For a simplicial complex $\mathcal{K}$, we define a filtration of $\mathcal{K}$ as a nested sequence of subcomplexes of $\mathcal{K}$, \begin{equation}\emptyset\subseteq\mathcal{K}_{0}\subseteq\mathcal{K}_{1}\ldots\subseteq\mathcal{K}_{n}=\mathcal{K}\end{equation}

In persistent homology, the nested sequence of subcomplexes usually depends on a filtration parameter. The persistence of a topological feature is denoted graphically by its life span with respect to filtration parameter. Subcomplexes corresponding to various filtration parameters offer the topological fingerprints over multiple scales. The $k^{th}$ persistence Betti number  $\beta^{i,j}_k$ is given by  the ranks of the $k^{th}$ homology groups of $\mathcal{K}_i$ that are alive and are defined as 
\begin{equation}\beta^{i,j}_k=\mbox{rank}(\mathcal{H}^{i,j}_k)=\mbox{rank}(\mathcal{Z}_k(\mathcal{K}_i)/(\mathcal{B}_k(\mathcal{K}_j)\cap\mathcal{Z}_k(\mathcal{K}_i))).
\end{equation}

\begin{figure}[h]
\centering
\begin{subfigure}{0.49\textwidth}
	\centering	
	\includegraphics[width=0.65\linewidth]{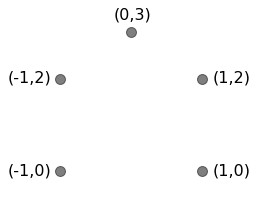}
	\caption{}
\end{subfigure}
\begin{subfigure}{0.49\textwidth}
	\centering	
	\includegraphics[width=1\linewidth]{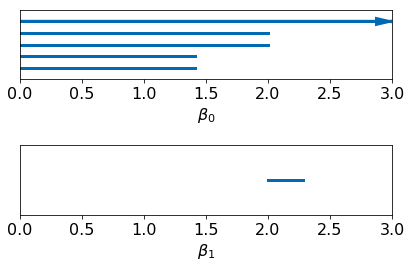}
	\caption{}
\end{subfigure}
\caption{(a) An example of 5 points in $\mathbb{R}^2$ and (b) the corresponding persistence barcodes. The length of each barcode corresponds to the persistence of each topological object ($\beta_0$,$\beta_1$,$\beta_2$,etc..) over the Vietoris-Rips (VR) complex filtration.}
\label{fig:barcode}
\end{figure}

The persistence of Betti numbers over the filtration interval can be recorded in many different ways. The commonly used ones are  persistence barcodes and persistence diagrams. An example of barcodes is provided in Figure \ref{fig:barcode}.

\subsubsection{Similarity and distance}
In this work, we  use Bottleneck and Wasserstein distances to extract atom-specific topological information and facilitate atom-specific persistent homology.  Let $X$ and $Y$ be multisets of data points, 
the  Bottleneck   and Wasserstein distances of  $X$ and $Y$ are given by  \cite{CEH07}
\begin{equation}
d_B(X,Y)=\displaystyle\inf_{\gamma\in B(X,Y)}\sup_{x\in X}\mid\mid x-\gamma(x)\mid\mid_{\infty},
\end{equation}
and \cite{cohen2010lipschitz}
\begin{equation}
d_W^p(X,Y)=\displaystyle\left(\inf_{\gamma\in B(X,Y)}\sum_{x\in X}\mid\mid x-\gamma(x)\mid\mid_{\infty}^p\right)^{1/p},
\end{equation}
respectively. Here  $B_(X,Y)$ is the collection of all bijections from $X$ to $Y$. Note that in our work, topological invariants of different dimensions are compared separately.

\subsubsection{Vietoris-Rips complex}
Given a metric space $M$ and a cutoff distance $d$, a simplex is formed if all points have pairwise distances no greater than $d$. All such simplices form the Vietoris-Rips (VR) complex. The abstract nature of the VR complex allows the construction of simplicial complexes from a correlation function, which models the pairwise interaction of atoms using a radial basis function versus more standard distance metrics.  {The R library TDA}  is used to generate persistence barcodes \cite{fasy2014introduction} . 

\subsubsection{Atom-specific persistent homology and element-specific persistent homology}\label{sect:PH_feat}

 Element-specific persistent homology was introduced to embed chemical and biology information into topological invariants \cite{ZXCang:2017a,ZXCang:2018a}. Its essential idea is to construct topological representations from subsets of atoms in various element types in a protein. For example, if one selects all carbon atoms in a protein, the resulting persistence barcodes will represent the strength and network of hydrophobicity in the protein.  

\begin{figure}[H]\centering
\includegraphics[width=0.8\textwidth]{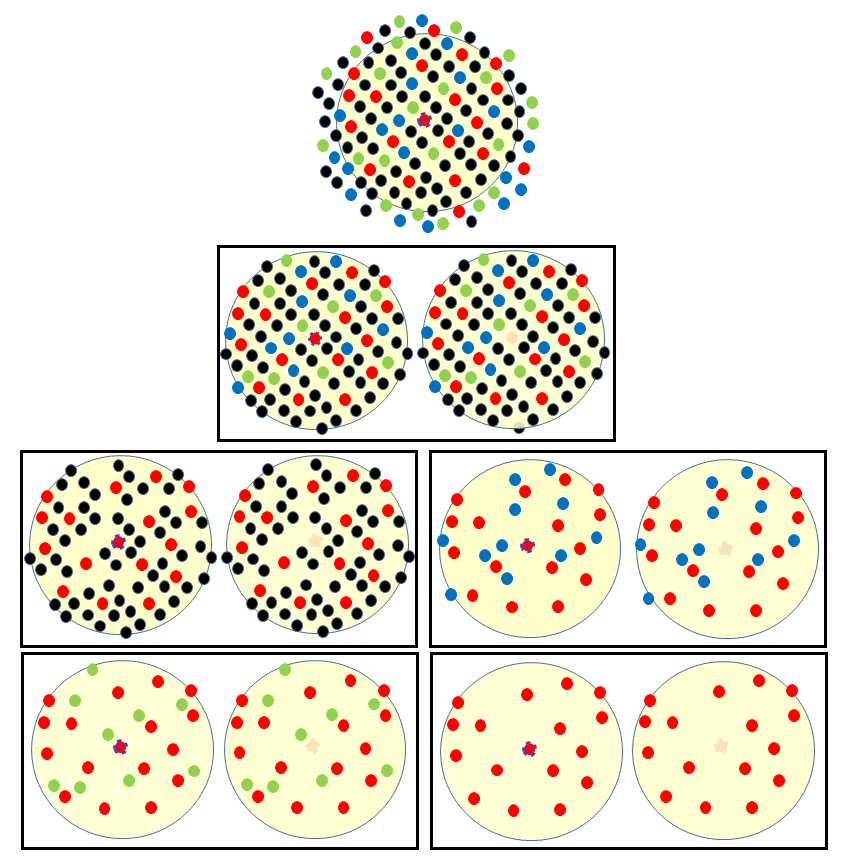}
\caption{Illustration of  Atom-specific persistent homology point clouds. Top: the original point cloud. The atom of interest is at the center of the circle. Second row: a pair of conjugated sets of point clouds for atom-specific persistent homology. The rest: Four pairs of conjugated point clouds for atom-specific and element-specific persistent homology.   } 
\label{fig:ASPH}
\end{figure}
In contrast, atom-specific persistent homology is designed to highlight the topological information of a given atom in a biomolecule. It creates two conjugated subsets of atoms centered around the atom of interest, one with and one without the specific atom.  Conjugated simplicial complexes, conjugated homology groups and conjugated topological invariants are generated for the conjugated sets of points clouds.  The difference between the conjugated topological invariants, measured by both Wasserstein and Bottleneck distances, offers a topological representation of the atom of interest.  As shown in Figure \ref{fig:ASPH}, atom-specific and element-specific conjugated point clouds can be constructed  for a given dataset. 

In this work, we focus on C$_{\alpha}$ B factor predictions. We use element specific persistent homology to enhance the topological representation of each  C$_{\alpha}$ neighborhood. Meanwhile, we develop atom-specific persistent homology to pinpoint the topological representation at each C$_{\alpha}$ atom.   With these selections of subsets, Vietoris-Rips complexes are constructed by contact maps or matrix filtration \cite{KLXia:2014c}. 

\begin{figure}[H]
\begin{subfigure}{0.39\textwidth}
\includegraphics[width=0.8\textwidth]{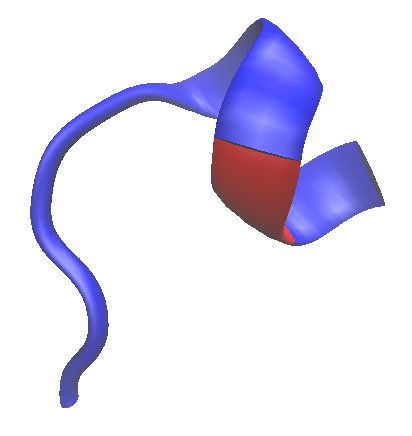}
\caption{}
\end{subfigure}
\begin{subfigure}{0.59\textwidth}
\includegraphics[width=1\textwidth]{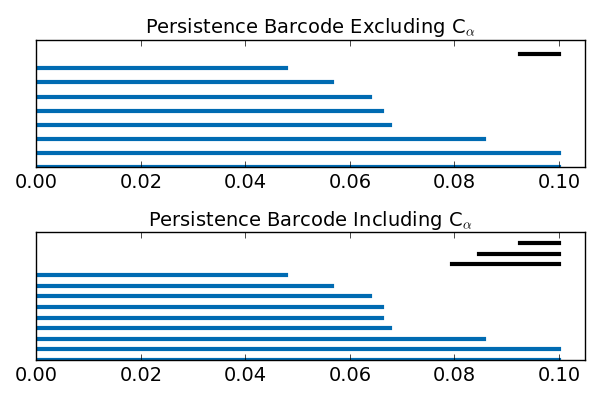}
\caption{}
\end{subfigure}
\caption{Illustration of residue 338 C$_\alpha$ atom-specific  persistent homology  in the ${\rm CC}$ element-specific point cloud of protein PDBID 1AIE. For this example residues 332-339 are used and are shown on the left. The C$_\alpha$ location used to generate the barcodes (right) is highlighted in red in the left chart. Conjugated persistence barcodes are generated with and without the selected C$_\alpha$. 
}
\label{fig:PH_gen}
\end{figure}

To capture element-specific interactions we consider three subsets of carbon-carbon, carbon-nitrogen, and carbon-oxygen point clouds. This gives us the following element specific pairs,
\begin{equation}
\mathcal{P}=\{{\rm CC,CN,CO}\}.
\end{equation}
For a given Protein Data Bank (PDB) file, persistence barcodes are calculated as follows. Given a specific C$_{\alpha}$ of interest, say ${\bf r}_i^k\in \mathcal{P}_k$ in an element specific set $\mathcal{P}_k$ ($\mathcal{P}_1={\rm CC}, \mathcal{P}_2={\rm CN}$, and  $\mathcal{P}_3={\rm CO}$) ,   a point cloud consisting of all atoms within a pre-defined cutoff radius $r_c$ is selected:  
\begin{equation}
\mathcal{R}^k_i= \{\mathbf{r}^{k}_{j}\bigm| ||\mathbf{r}^{k}_{i}-\mathbf{r}^{k}_{j}||<r_c, \quad {\bf r}_i^k, {\bf r}_j^k\in \mathcal{P}_k, \forall\ j\in 1,2,\ldots N\},
\end{equation}
where $N$ is the number of atoms in the $k$th element pair $\mathcal{P}_k$. A conjugated set of point cloud,  $\hat{\mathcal{R}}^k_i$, includes the same set of atoms, except for  ${\bf r}_i^k$.    For a given pair of conjugated point clouds  $\mathcal{R}^k_i$ and  $\hat{\mathcal{R}}^k_i$, conjugated  simplicial complexes, conjugated homology groups,  and conjugated persistence barcodes  are computed via persistent homology.  We compute  Euclidean  distance based filtration using the Vietoris-Rips complex. Additionally, for a given set of atoms selected according to atom-specific and element specific constructions,    we generate a  family of multiresolution persistence barcodes by a resolution controlled filtration matrix:    \cite{KLXia:2014c}
\begin{equation}
M_{nm}(\vartheta )= 1-\Phi(||\mathbf{r}_n-\mathbf{r}_m||;\vartheta),
\end{equation}
where $\vartheta$ denotes a set of kernel parameters. We have used  both exponential kernels  
\begin{equation}
\Phi(||\mathbf{r}_n-\mathbf{r}_m||; \eta, \kappa )=\displaystyle{e^{-(||\mathbf{r}_n-\mathbf{r}_{m}||/ \eta)}}^{\kappa},\qquad \kappa>0
\end{equation}
and  Lorentz kernels  
\begin{equation}
\Phi(||\mathbf{r}_n-\mathbf{r}_m||;\eta, \nu)=\dfrac{1}{1+\big(||\mathbf{r}_n-\mathbf{r}_{m}||/\eta \big)^{\nu}}, \qquad \nu>0
\end{equation}
where $\eta$ $\kappa$, and $\nu$ are pre-defined constants. This filtration matrix is used in association with the Vietoris-Rips complex to generate persistence barcodes or persistence diagrams. Then these topological invariants are  compared using both Bottleneck and Wasserstein distances. An example of the conjugated persistence barcode pair generated for a C$_\alpha$ atom is illustrated in Figure \ref{fig:PH_gen}.

\subsection{Machine learning models}\label{ML_Models}

Topological  features are used for prediction of protein B factor using both least squares fitting and machine learning as described in the following subsections.

\subsubsection{Gradient boosted trees}
Gradient boosting is an ensemble method that uses a number of  ``weak learners'' to construct a prediction model in an iterative manner. The method is optimized via   gradient descent, which minimizes the residuals of  a loss function. At each step of the gradient boosting, gradient boosting trees (GBTs) incorporate decision trees to improve their predictive power. Ensemble methods like GBTs are useful because they can handle  a diverse feature set, have strong predictive power, and are typically robust to outliers and against overfitting.

In this work, we optimize the GBT hyper-parameters using the standard practice of a grid search. The parameters used for testing are provided in Table \ref{tab:GBT_parameters}. Any hyper-parameters not listed in the table were taken to be the default values provided by the python scikit-learn package.

\begin{table}[H]
\centering
\begin{tabular}{l|l} \hline \hline
Parameter 	   	  & Setting \\ \hline
Loss Function	  & Quantile\\ 
Alpha			  & 0.975\\ 
Estimators 	 	  & 500\\ 
Learning Rate 	  & 0.25\\
Max Depth 		  & 4\\ 
Min Samples Leaf  & 9\\ 
Min Samples Split & 9\\ \hline\hline
\end{tabular}
\caption{Boosted gradient tree hyperparameters used for testing. Parameters were determined using a grid search. Any hyperparameters that is not listed were taken to be the default values provided by the python scikit-learn package.}
\label{tab:GBT_parameters}
\end{table}

\subsubsection{Deep learning with a convolutional neural network}
Neural networks are modeled after the function of neurons in brain. A neural network applies activation functions, called perceptrons, to  inputs. Weights of the network are trained to minimize a loss function over many epochs, or passes of an entire training dataset. When a neural network has several layers of perceptrons we call it a deep neural network (DNN) and the intermediate  layers are known as hidden layers.

\begin{figure}[H]
\centering
\includegraphics[width=1\textwidth]{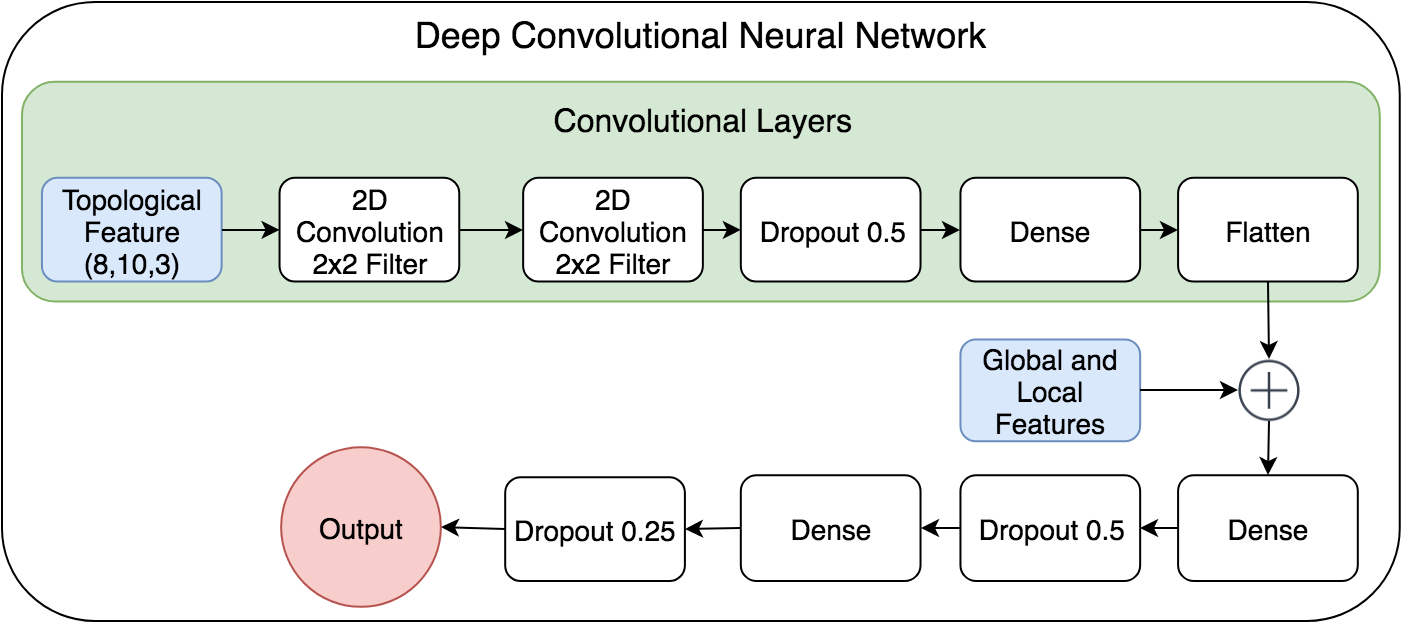}
\caption{The deep learning architecture using a convolutional neural network combined with a deep neural network. The plus symbol represents the concatenation of features.}
\label{fig:CNN}
\end{figure}

Convolutional neural networks (CNNs) have recently had great success in image classification. Using convolutions of a pre-defined filter size and number of filters,  CNNs can automatically extract high-level features from input images. CNNs are advantageous because they can perform as well as other models without training as many parameters as a densely connected deep neural network. By applying several convolutions one can extract high-level features of an image. In this work we generate a image-like heat map by using a range of kernel parameters for atom-specific and element-specific  persistent homology. The CNN output is then flattened and fed as input to a DNN along with global and local protein features. This allows us to use the same feature set as the boosted gradient method as well as the generated PH image data. A diagram of the CNN architecture is provided in Figure \ref{fig:CNN}.

For each C$_{\alpha}$ of the training set, the CNN is passed a three-channel persistent homology image of dimension (8,10,3). The model takes the input image data and applies two convolutional layers with 2x2 filters followed by a dropout of 0.5. The image data is passed through a dense layer, flattened, then joined with the other global and local features to form a dense layer of 218 neurons. This is followed by a dropout layer of 0.5, another dense layer of 100 neurons, a dropout layer of 0.25, a dense layer of 10 neurons, and finishes with a dense layer of  {output}. Figure \ref{fig:CNN} provides an illustration   of the deep CNN used in this work.

The deep convolutional neural network has several hyper-parameters that can be tuned. As with the GBT, the deep convolutional neural network hyper-parameters are optimized using a basic grid search. Table \ref{tab:param_CNN} provides the parameters used for testing. Any hyper-parameters that are not listed below were taken to be the default values provided by the python Keras package.

\begin{table}[H]
\centering
\caption{Convolutional Neural Network (CNN) parameters used for testing. Parameters were determined using a grid search. Any hyper-parameters not listed below were taken to be the default values provided by python with the Keras package.}
\begin{tabular}{ll} \hline\hline
Parameter 	   	  & Setting \\ \hline
Learning Rate 	  & 0.001\\
Epoch 	 	  	  & 1000\\
Batch Size		  & 1000\\
Loss			  & Mean Squared Error\\ 
Optimizer		  & Adam \\ \hline\hline
\end{tabular}
\label{tab:param_CNN}
\end{table}

\subsubsection{Consensus method}

 In this work, we combine the predictions of two machine learning models to construct a simple consensus model. The consensus prediction used in this work is generated by the average of  C$_{\alpha}$ B factor values predicted from the GBT and deep CNN models.

\subsection{Machine learning features}\label{sec:ML_features}

A variety of element-specific and atom-specific persistence barcodes were generated using the techniques discussed in Sec. \ref{sect:PH_feat}. In this work, we include 60 topological features. These features are generated in several ways by varying: kernels (Lorentz and exponential), element-specific pairs (CC, CN, CO),   and  distance metrics (Wasserstein-0 and Wasserstein-1, Bottleneck-0 and Bottleneck-1). For this work all  persistent homology features were generated with the cutoff of 11\AA.

\subsubsection{  Wasserstein and Bottleneck metrics for modified persistence diagrams}

\begin{figure}[H]
\centering
\begin{subfigure}[t]{0.325\textwidth}
\includegraphics[width=1\textwidth]{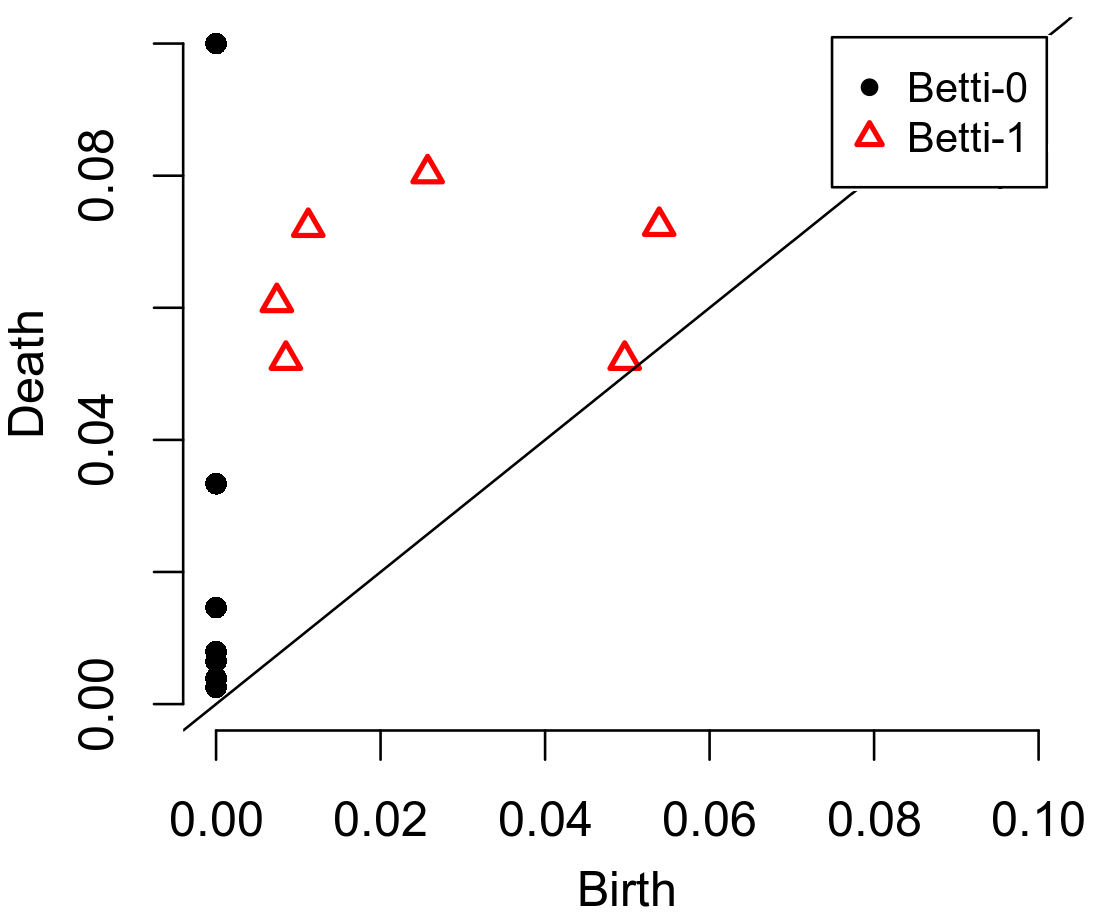}
\end{subfigure}
\begin{subfigure}[t]{0.325\textwidth}
\includegraphics[width=1\textwidth]{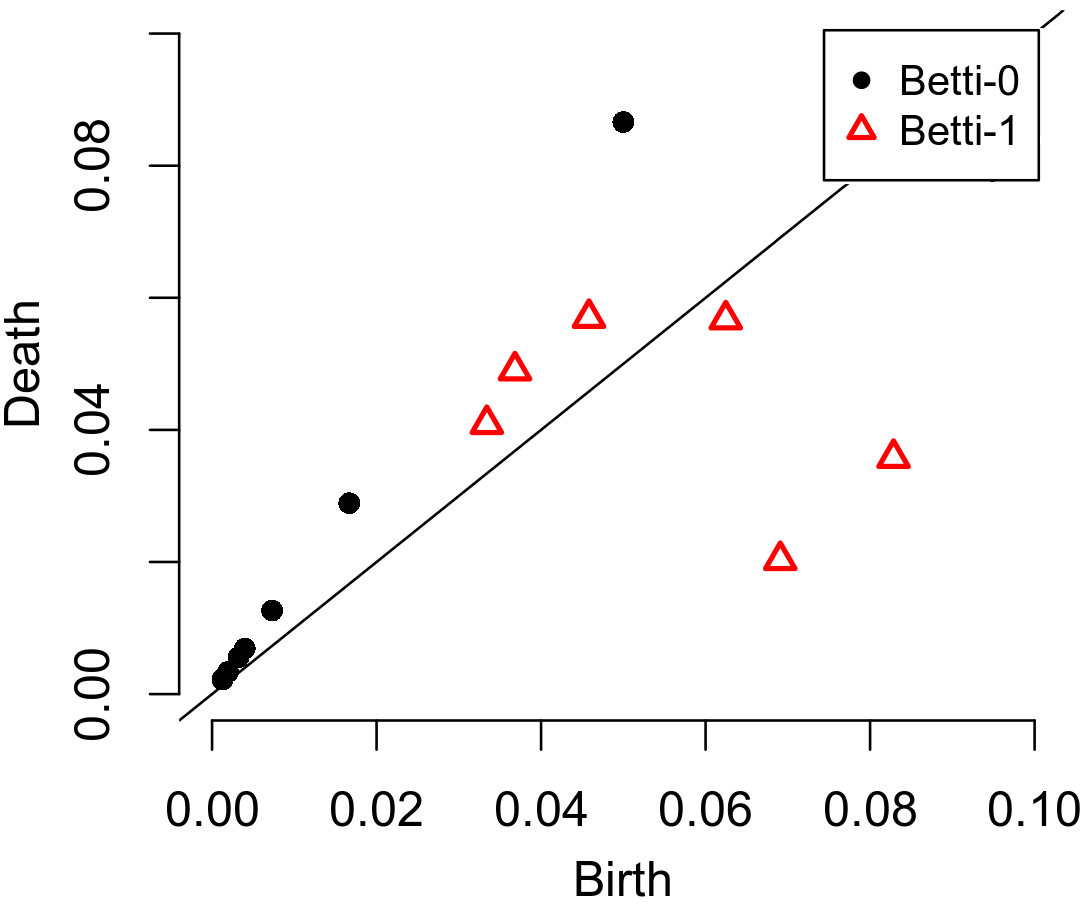}
\end{subfigure}
\begin{subfigure}[t]{0.325\textwidth}
\includegraphics[width=1\textwidth]{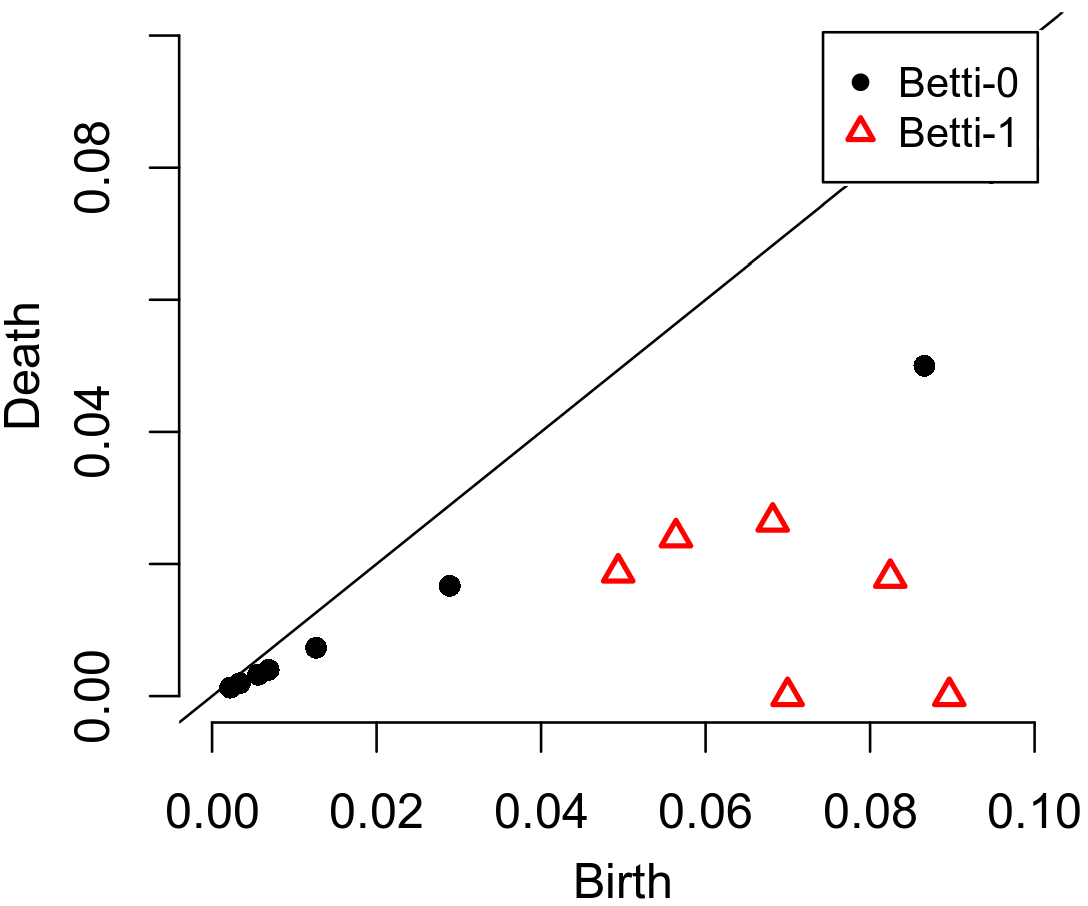}
\end{subfigure}
\caption{Illustration of modified persistence diagrams used  in distance calculations.  Left:  Unchanged. Middle:  Rotated $30^{\circ}$. Right:  rotated $60^{\circ}$. Black dots are Betti-0 events and triangles  are Betti-1 events.}
\label{fig:Rips_Mod}
\end{figure}

The distances evaluated from  Wasserstein and Bottleneck evaluations of persistence diagrams depend on the boundary of the diagrams. Specifically, when two persistence diagrams are compared, the extra events on one diagram that do not match any events on the other diagram might contribute to the final distance by their distances from the boundary. For this reason,  
we create two additional persistence diagrams in which the $y$-axis is rotated clockwise by  $30^{\circ}$ or $60^{\circ}$, respectively, see Figure \ref{fig:Rips_Mod}.  This modification changes the Bottleneck and Wasserstein distances and allows the model to recognize elements that have a short persistence (i.e. have a short lifespan). Lastly, we modified the persistence diagram by  reflecting around the diagonal axis. An example of this modification is illustrated in Figure \ref{fig:Rips_Mod}.  Table \ref{tab:ML_feats} provides  a list of  kernels, kernel parameters, $y$-axis change,  distance metric, and element-specific pairs used to generate features in machine learning models.

\begin{table}[H]
\centering
\begin{tabular}{cccccc} \hline\hline
No. features & Kernel & Kernel parameter & Diagram  & Distance metric & Element-specific pair\\ \hline
12                  & Lor & $\eta=21$, $\nu=5$ & Unchanged  & B, W &CC, CN, CO\\
12                  & Exp & $\eta=10$, $\kappa=1$ & Unchanged & B, W &CC, CN, CO\\
12                  & Exp & $\eta=2$, $\kappa=1$  &  Diagonal reflection  & B, W &CC, CN, CO\\
12                  & Exp & $\eta=2$, $\kappa=1$  & Rotated $30^{\circ}$ & B, W &CC, CN, CO\\
12                  & Exp & $\eta=2$, $\kappa=1$  & Rotated $60^{\circ}$ & B, W &CC, CN, CO\\ \hline\hline
\end{tabular}
\caption{Parameters used for topological feature generation. All features used a cutoff of 11\AA. Both lorentz (Lor) and exponential (exp) kernels and Bottleneck (B) and Wasserstein (W) distance metrics were used.}
\label{tab:ML_feats}
\end{table}
Other features include global features  {from} PDB files, i.e., R-value, protein resolution, and number of heavy atoms. Additional local features include packing density, amino acid type, occupancy, and secondary structure information generated by STRIDE software \cite{heinig2004stride}. 

\subsubsection{Image-like  persistent homology features}
\label{sec:PH_image}
Using the process described in Section \ref{sect:PH_feat} we generate 2D image-like  persistent homology   features, $F^k_i=\{f^k_i(\eta,\kappa) \}$, for each C$_{\alpha}$ of the proteins in the dataset by using various values of $\eta$ and $\kappa$ in the kernel function. A cutoff of 11 \AA\ with an exponential kernel and different values of $\eta$ and $\kappa$ are used to capture a wide variety of scales. In particular we use \[\eta=\{1,2,3,4,5,10,15,20\},\] and \[\kappa=\{1,2,3,4,5,6,7,8,9,10\}.\] 
The image-like matrix is given by ${ F}_i^k$ in Eq. (\ref{mat:1}), where each atom ${F}_i^k $ represents the PH feature of the $i^{th}$ C$_\alpha$ atom, and $k^{th}$ atom interaction (C, N, or O). 

\newcommand{\nm}[1]{f_i^k(#1)}
\newcommand\undermat[2]{%
  \makebox[0pt][l]{$\smash{\underbrace{\phantom{%
    \begin{matrix}#2\end{matrix}}}_{\text{$#1$}}}$}#2}
{    \small
\begin{equation}\left. {F}_i^k = \begin{bmatrix}
\nm{1,1} & \nm{1,2} & \ldots & \nm{1,9} & \nm{1,10} \\
\nm{2,1} & \nm{2,2} & \ldots & \nm{2,9}& \nm{2,10} \\
\vdots 	  &	 			&		 \vdots \\
\nm{15,1} & \nm{15,2} & \ldots & \nm{15,9}& \nm{15,10}\\
\undermat{\kappa}{\nm{20,1} & \nm{20,2} & \ldots & \nm{20,9}& \nm{20,10} } 
\end{bmatrix}\right\}\eta
\label{mat:1}
\end{equation}
}\ \\

This results in 2D PH images of dimension (8,10). Images are created for element-specific  C$_{\alpha}$ interactions with carbon, nitrogen, and oxygen atom giving each image three channels. This results in a final image dimension of (8,10,3) for each C$_\alpha$ atom.

%

\section{Results}
\label{sect:Num}

\subsection{Data sets}
In this work, we use two data sets, one from Refs. \cite{Opron:2014,Opron:2015a} and the other from Park, Jernigan, and Wu  \cite{JKPark:2013}. The first contains 364 proteins \cite{Opron:2014,Opron:2015a} and the second contains 3 subsets of small, medium, and large proteins  \cite{JKPark:2013}. All sequences have a resolution of 3 {\AA} or higher and an average resolution of 1.3  {\AA}  and the sets include proteins that range from 4 to 3912 residues \cite{JKPark:2013}.

For all testing, we exclude protein 1AGN due to known problems with this protein data \cite{Opron:2015a}. Proteins 1NKO, 2OCT, and 3FVA are also excluded because these proteins have residues with B factors reported as zero, which is unphysical. For the machine learning results, proteins 1OB4, 1OB7, 2OLX, and 3MD5 are excluded because the STRIDE software is unable to provide secondary features for these proteins. The image like features used in all convolutional neural networks were standardized with mean 0 and variance of 1

\subsection{Evaluation metric} \label{kernels}

  We use the proposed methods to predict B factors of all C$_{\alpha}$ atoms present in a protein. Linear least square fitting was done using  {only} topological features. The machine learning models were executed using a leave-one-(protein)-out method to blindly predict the B factors of all C$_{\alpha}$ atoms in each protein. The machine learning models were trained using the data and features described in Sections \ref{sect:PH_feat}, \ref{ML_Models}, \ref{sec:ML_features}. For comparison, we include previously existing C$_{\alpha}$ B factor prediction fitting methods.

To quantitatively assess our method for B factor prediction we use the Pearson correlation coefficient  given by 
\begin{equation}
{\rm PCC}=\dfrac{\ds\sum_{i=1}^{N}(B_i^e-\bar{B}^e)(B_i^t-\bar{B}^t)}{\bigg[\ds\sum_{i=1}^{N}(B_i^e-\bar{B}^e)^2\ds\sum_{i=1}^{N}(B_i^t-\bar{B}^t)^2\bigg]^{1/2}},
\end{equation}
where $B^t_i,i = 1, 2,\ldots,N$ 
are  predicted B factors using the proposed method and $B^e_i ,i = 1, 2,\ldots,N$ experimental B factors from the PDB file. The terms $B^t_i$ and $B^e_i$ represent the $i^{th}$ theoretical and experimental B factors respectively.
Here $\bar{B}^e$ and  $\bar{B}^t$ are averaged B factors.  

\normalsize

\subsection{Cutoff distance}

\begin{table}[H]
\centering
\begin{tabular}{lccc}\hline\hline
Kernel Type 	& $\nu$ & $\eta^n$ & $\kappa$ \\ \hline 
Lorentz ($n=1$) 	&5	&21	   &-\\
Exponential ($n=2$)  & - &  10  & 1 \\ \hline \hline

\end{tabular}
\caption{Parameters used for the persistent homology element specific features with a cutoff of 11 \AA.}
\label{tab:par}
\end{table}

\begin{figure}[H]
\centering
\includegraphics[width=0.5\textwidth]{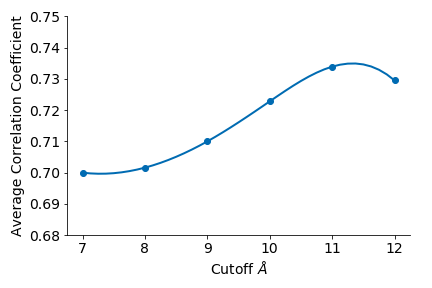}
\caption{Average pearson correlation coefficient over the entire protein dataset fitting all 24 persistent homology features using various cuttoff distances.}
\label{fig:cut}
\end{figure}

In this work, the optimal cutoff of $r_c =11$\AA\ is found over a grid search using various cutoff distances. Figure \ref{fig:cut} displays the average Pearson correlation coefficient, obtained via fitting, over an entire dataset of 364 protein using all persistent homology metrics with various point cloud distance cutoffs.
For each protein we use the parameters listed in Table \ref{tab:par}. 
The values used in this work were determined using the standard practice of a grid search.

\subsection{Least squares fitting within proteins}

\begin{table}{H}
\centering
\caption{Average Pearson correlation coefficients of least squares fitting C$_{\alpha}$ B factor prediction of small, medium, large, and superset using 11\AA\ cutoff. Two Bottleneck (B) and Wasserstein (W) metrics using various kernel choices are included. Results for pFRI are taken from Opron et al\cite{Opron:2014}. GNM and NMA value are taken from the course grained C$_{\alpha}$ results reported in Park {\it et al} \cite{JKPark:2013}.}
\label{tab:avg}
\begin{tabular}{c|ccc|ccc|ccc|ccc}\hline\hline
& \multicolumn{3}{|c|}{B \& W}& \multicolumn{3}{|c|}{B} 	& \multicolumn{3}{|c|}{W}&&\\ \cline{2-10}
&	Exp	&	Lor	&	Both	&	Exp	&	Lor	&	Both	&	Exp	&	Lor	&	Both	&	pFRI	&	GNM	&	NMA	\\ \hline
Small&0.87&0.84&0.94&0.74&0.72&0.85&0.74&0.73&0.86&0.59&0.54&0.48\\
Medium&0.68&0.68&0.78&0.62&0.61&0.69&0.60&0.63&0.69&0.61&0.55&0.48\\
Large &0.61&0.60&0.70&0.54&0.54&0.61&0.51&0.55&0.62&0.59&0.53&0.49\\
Superset&0.65&0.64&0.73&0.58&0.58&0.65&0.55&0.59&0.66&0.63&0.57&NA\\ \hline\hline
\end{tabular}
\end{table}

The Pearson correlation coefficients using least squares fitting for C$_{\alpha}$ B factor prediction of small, medium, and large protein subsets are provided in Tables \ref{tab:small}, \ref{tab:med}, and \ref{tab:lar} respectively. Results for the all proteins in the dataset are provided in Table \ref{tab:all}. The average Pearson correlation coefficients for small, medium, large, and superset data sets are provided in Table \ref{tab:avg}. Table \ref{tab:avg} includes fitting results using only Bottleneck, only Wasserstein, and using both Bottleneck and Wasserstein metrics. We also include results using only exponential kernel, only a Lorentz kernel, or both an exponential and Lorentz kernel for fitting. All results reported here PH features generated with a cutoff of 11\AA\ and include three element-specific subsets (carbon-carbon, carbon-nitrogen, carbon-oxygen). Overall fitting methods using the various persistent homology features performed similarly. The best results came from using features generated by both exponential and Lorentz kernels and both Bottleneck and Wasserstein distances. Using both kernels and both distance metrics resulted in an average correlation coefficient of 0.73 for the superset.

\subsection{Blind machine learning prediction across proteins}

The aforementioned  least  squares   fitting methods cannot predict the B factors of unknown proteins.  Machine learning methods enable us to blindly predict B factors across proteins. In this section, we utilize  {both} boosted gradient  {and} convolutional neural network  {algorithms} for the blind prediction of B factor across different proteins.  {Taken together, the entire dataset contains}  more  than  620  000  atoms.  {We use a leave-one-protein out cross validation in our prediction. That is, for each protein, the data from a protein whose B factors will be predicted, is excluded from the training data}. This gives rise to  a training  set  of  roughly  600  000  data  points  {for each protein} (i.e., atoms and associated B factors). The Pearson correlation coefficients using boosted gradient (GBT), convolutional neural network (CNN), and consensus method (CON) for C$_{\alpha}$ B factor prediction of small, medium, and large protein subsets are provided in Tables \ref{tab:Small_ML}, \ref{tab:Medium_ML}, and \ref{tab:Large_ML} respectively. Parameters for GBT and CNN methods can be found in Tables \ref{tab:GBT_parameters} and \ref{tab:param_CNN}. The global and local features used for training and testing are provided in Section \ref{sec:ML_features}. Results for all proteins are provided in Table \ref{tab:365_ML}. The average Pearson correlation coefficients for small, medium, large, and superset data sets are  provided in Table \ref{tab:Avg_ML}. All results reported here use a cutoff of 11\AA\ and include three element-specific subsets (carbon-carbon, carbon-nitrogen, carbon-oxygen). Kernel parameters for both exponential and Lorentz kernels are provided in Table \ref{tab:par}. Results from previously existing C$_{\alpha}$ B factor prediction methods are included for comparison in Table \ref{tab:Avg_ML}. Overall both GBT and CNN algorithms perform similarly. As expected, the CNN method outperforms the GBT with average correlation coefficients over the superset of 0.60 and 0.59, respectively. The consensus method improves upon both results with an average Pearson correlation coefficient of 0.61 over the superset. Table \ref{tab:Avg_ML} shows that the blind prediction machine learning models perform better than fitting models GNM and NMA and similar to the pFRI fitting model. 
\begin{table}
\centering
\caption{Average Pearson correlation coefficients C$_\alpha$ B factor predictions for small-, medium-, and large-sized protein sets along with the entire superset of the 364 protein dataset. Gradient boosted tree (GBT), convolutional neural network, and consensus (CON) results are obtained by leave-one-protein-out (blind). The results of parameter-free flexibility-rigidity index (pfFRI), Gaussian network model (GNM) and normal mode analysis (NMA) were obtained via the least squares fitting of individual proteins.}
\label{tab:Avg_ML}
\begin{tabular}{*{8}{c}}\hline\hline
	&	CNN	&	GBT	&	CON	&	pFRI	&	GNM	&	NMA	\\	\hline
Small	&	0.63	&	0.58	&	0.62	&	0.59	&	0.54	&	0.48	\\	
Medium	&	0.60	&	0.58	&	0.61	&	0.61	&	0.55	&	0.48	\\	
Large	&	0.58	&	0.59	&	0.58	&	0.59	&	0.53	&	0.49	\\	
Superset	&	0.60	&	0.59	&	0.61	&	0.63	&	0.57	&	NA	\\ \hline\hline	
\end{tabular}
\end{table}

\section{Conclusion}\label{sect:con}

An essential component of the paradigm of protein dynamics is the correlation between protein flexibility and protein function. The shear complexity and large number of degrees of freedom make quantitative understanding of flexibility and function an inherently difficult problem. Several time-independent methods for predicting protein B factors exist. Examples include NMA\cite{Brooks:1983,Go:1983,Levitt:1985,Tasumi:1982}, ENM \cite{Tirion:1996}, GNM \cite{Bahar:1997,Bahar:1998,Brooks:1983harmonic}, and FRI methods \cite{KLXia:2013f, Opron:2014,Opron:2015a,Opron:2016a}. None of the methods above are able to blindly predict protein B factors of an unknown protein. We hypothesize that the intrinsic physics of proteins lie in a low-dimensional space embedded in a high-dimensional data space. Based on this hypothesis the authors previously introduced the graph theory based multiscale weighted colored graph (MWCG) \cite{DBramer:2018a,DBramer:2018b}. The authors showed that MWCG's are able to  successfully blindly predict cross-protein B factors. 

In this work we explore this hypothesis further by creating a B factor predictor using tools from algebraic topology. In order to construct localized topological representations for individual atoms from global topological tools, we propose atom-specific topology    {and} atom-specific persistent homology. This approach creates two conjugated sets of atoms: the first set is centered around the given atom of interest while the other set is  {identical} but excludes the atom of interest. Element-specific selections are further implemented to embed biological information into atom-specific persistent homology.  
The distance between the topological invariants generated from these conjugated sets of atoms is used to represent the atom of interest. Both   Bottleneck and Wasserstein metrics are utilized to estimate the topological distances between conjugated barcodes. The Vietoris-Rips complex is employed for topological barcode generation.

To test the proposed method we use over 300 proteins or more than 600,000 B factors. Atom-specific persistent homology features are generated using several element-specific interactions, kernel choices, parametrizations, and barcode distance metrics. First we employ topological features to fit protein B factors using linear least squares. Using topological features our fitting model outperformed previous fitting models with an average Pearson correlation coefficient of 0.73 over the superset of proteins. Next we considered using the topological features to blindly predict protein B factors of C$_\alpha$ atoms. We generated two machine learning models, a gradient boosted tree (GBT) and deep convolutional neural network (CNN). Additionally we averaged the C$_\alpha$ prediction from the two models to generate a more robust consensus model. A variety of local and global features were included in addition to the generated topological features. Our blind prediction consensus model outperformed both GNM and NMA fitting models and produced similar results to the pFRI fitting model.

To the authors' knowledge, this work is the first time persistent homology has been used to predict the B factor of atoms in a protein. This approach is novel because topology is a global property and on its own cannot be used to describe  local atomic information. Our unique approach allows us to create local topological representation  with a variety of customizable parameters using a global mathematical tool. This allows the model to account for multiple spatial interaction scales and element specific interactions. Our results demonstrate that this is a accurate and robust approach. Moreover, the results could easily be improved by including a larger dataset, fine tuning parameters, and exploring different machine learning approaches.

This method can be applied to a variety of interesting applications related to protein dynamics. Examples include  allosteric site detection, computer-aided drug design, hinge detection, hot spot identification,   and protein folding stability changes upon mutation. More generally this method may be amenable to problems outside proteins such as network dynamics and social network centrality measure.

\section*{Acknowledgment} 

This work was supported in part by  NSF Grants DMS-1721024 and DMS-1761320, and NIH grant  GM126189.  

\section{Appendix}
\fontsize{10}{10}
\begin{table}[H]
\centering
\caption{Pearson correlation coefficients for cross protein  C$_{\alpha}$ atom blind B factor prediction obtained by boosted gradient (GBT), convolutional neural network (CNN), and consensus (CON) for the small-sized protein set.\\
\label{tab:Small_ML}}
\begin{tabular}{ccccc}\hline\hline 
PDB ID&	$N$&	GBT&	CNN&	CON \\ \hline
1AIE&	31&	0.75&	0.7&	0.78\\
1AKG&	16&	0.27&	0.32&	0.29\\
1BX7&	51&	0.74&	0.74&	0.76\\
1ETL&	12&	0.37&	0.82&	0.55\\
1ETM&	12&	0.37&	0.63&	0.43\\
1ETN&	12&	0.07&	0.48&	0.13\\
1FF4&	65&	0.61&	0.66&	0.64\\
1GK7&	39&	0.77&	0.9&	0.82\\
1GVD&	56&	0.71&	0.55&	0.69\\
1HJE&	13&	0.84&	0.75&	0.9\\
1KYC&	15&	0.62&	0.69&	0.66\\
1NOT&	13&	0.69&	0.96&	0.8\\
1O06&	22&	0.94&	0.93&	0.95\\
1P9I&	29&	0.73&	0.73&	0.74\\
1PEF&	18&	0.79&	0.82&	0.82\\
1PEN&	16&	0.36&	0.74&	0.44\\
1Q9B&	44&	0.59&	0.85&	0.67\\
1RJU&	36&	0.6&	0.46&	0.58\\
1U06&	55&	0.44&	0.4&	0.45\\
1UOY&	64&	0.72&	0.7&	0.76\\
1USE&	47&	0.05&	0.32&	0.12\\
1VRZ&	13&	0.54&	0.34&	0.54\\
1XY2&	8&	0.79&	0.82&	0.81\\
1YJO&	6&	0.7&	-0.06&	0.57\\
1YZM&	46&	0.69&	0.64&	0.7\\
2DSX&	52&	0.34&	0.34&	0.36\\
2JKU&	38&	0.57&	0.71&	0.66\\
2NLS&	36&	0.23&	0.47&	0.29\\
2OL9&	6&	0.94&	0.85&	0.94\\
6RXN&	45&	0.59&	0.6&	0.61\\ \hline\hline
\end{tabular}
\end{table}

\begin{table}[H]
\centering
\begin{tabular}{ccccc}
\caption{Pearson correlation coefficients for cross protein C$_{\alpha}$ atom blind B factor prediction obtained by boosted gradient (GBT), convolutional neural network (CNN), and consensus (CON) for the medium-sized protein set.\label{tab:Medium_ML}} \\
\hline\hline 
PDB ID&	$N$&	GBT&	CNN&	CON \\ \hline
1ABA&	87&	0.73&	0.71&	0.74\\
1CYO&	88&	0.64&	0.7&	0.68\\
1FK5&	93&	0.59&	0.6&	0.61\\
1GXU&	89&	0.67&	0.68&	0.69\\
1I71&	83&	0.53&	0.58&	0.56\\
1LR7&	73&	0.62&	0.61&	0.64\\
1N7E&	95&	0.63&	0.58&	0.65\\
1NNX&	93&	0.78&	0.79&	0.8\\
1NOA&	113&0.55&	0.53&	0.56\\
1OPD&	85&	0.42&	0.34&	0.41\\
1QAU&	112&0.51&	0.59&	0.57\\
1R7J&	90&	0.71&	0.77&	0.75\\
1UHA&	82&	0.71&	0.74&	0.73\\
1ULR&	87&	0.54&	0.53&	0.56\\
1USM&	77&	0.73&	0.72&	0.75\\
1V05&	96&	0.6&	0.64&	0.63\\
1W2L&	97&	0.43&	0.5&	0.47\\
1X3O&	80&	0.41&	0.43&	0.44\\
1Z21&	96&	0.68&	0.65&	0.69\\
1ZVA&	75&	0.7&	0.7&	0.71\\
2BF9&	35&	0.48&	0.79&	0.58\\
2BRF&	103&0.72&	0.77&	0.75\\
2CE0&	109&0.6&	0.66&	0.64\\
2E3H&	81&	0.65&	0.68&	0.67\\
2EAQ&	89&	0.57&	0.63&	0.61\\
2EHS&	75&	0.62&	0.67&	0.65\\
2FQ3&	85&	0.77&	0.82&	0.81\\
2IP6&	87&	0.6&	0.66&	0.63\\
2MCM&	112&0.71&	0.77&	0.75\\
2NUH&	104&0.72&	0.56&	0.7\\
2PKT&	93&	0.01&	-0.04&	-0.01\\
2PLT&	98&	0.52&	0.53&	0.54\\
2QJL&	107&0.54&	0.57&	0.56\\
2RB8&	93&	0.67&	0.7&	0.7\\
3BZQ&	99&	0.45&	0.53&	0.49\\
5CYT&	103&0.39&	0.34&	0.39\\ \hline \hline

\end{tabular}
\end{table}

\begin{table}[H]
\centering
\begin{tabular}{ccccc}
\caption{Pearson correlation coefficients for cross protein C$_{\alpha}$ atom blind B factor prediction obtained boosted gradient (GBT), convolutional neural network (CNN), and consensus (CON) for the large-sized protein set.
\label{tab:Large_ML}}  \\
\hline\hline 
PDB ID&	$N$&	GBT&	CNN&	CON \\ \hline
1AHO&	66&	0.66&	0.66&	0.7\\
1ATG&	231&	0.55&	0.51&	0.55\\
1BYI&	238&	0.61&	0.5&	0.6\\
1CCR&	109&	0.55&	0.6&	0.59\\
1E5K&	188&	0.74&	0.72&	0.74\\
1EW4&	106&	0.59&	0.6&	0.61\\
1IFR&	113&	0.7&	0.64&	0.7\\
1NLS&	238&	0.55&	0.57&	0.57\\
1O08&	221&	0.49&	0.47&	0.49\\
1PMY&	123&	0.59&	0.7&	0.65\\
1PZ4&	113&	0.72&	0.8&	0.77\\
1QTO&	122&	0.53&	0.48&	0.54\\
1RRO&	108&	0.4&	0.45&	0.43\\
1UKU&	102&	0.75&	0.76&	0.77\\
1V70&	105&	0.63&	0.62&	0.64\\
1WBE&	206&	0.6&	0.56&	0.6\\
1WHI&	122&	0.59&	0.56&	0.6\\
1WPA&	107&	0.65&	0.65&	0.67\\
2AGK&	233&	0.67&	0.63&	0.67\\
2C71&	225&	0.57&	0.6&	0.6\\
2CG7&	110&	0.3&	0.32&	0.32\\
2CWS&	235&	0.61&	0.47&	0.6\\
2HQK&	232&	0.77&	0.77&	0.78\\
2HYK&	237&	0.65&	0.63&	0.65\\
2I24&	113&	0.44&	0.46&	0.46\\
2IMF&	203&	0.53&	0.58&	0.56\\
2PPN&	122&	0.64&	0.54&	0.63\\
2R16&	185&	0.44&	0.49&	0.46\\
2V9V&	149&	0.53&	0.52&	0.54\\
2VIM&	114&	0.44&	0.47&	0.47\\
2VPA&	217&	0.66&	0.75&	0.71\\
2VYO&	207&	0.6&	0.63&	0.63\\
3SEB&	238&	0.63&	0.6&	0.63\\
3VUB&	101&	0.59&	0.55&	0.59\\ \hline\hline

\end{tabular}
\end{table}

{\normalsize
\begin{longtable}[H]{llccc|llccc}
\caption{Pearson correlation coefficients for cross protein C$_{\alpha}$ atom blind B factor prediction obtained by boosted gradient (GBT), convolutional neural network (CNN), and consensus method (CON) for the Superset.\\ } \label{tab:365_ML} \\
\hline\hline \multicolumn{1}{c}{\textbf{PDB ID}} & \multicolumn{1}{c}{$N$} & \multicolumn{1}{c}{\textbf{GBT}} & \multicolumn{1}{c}{\textbf{CNN}} &  \multicolumn{1}{c|}{\textbf{CON}} &
\multicolumn{1}{c}{\textbf{PDB ID}} & \multicolumn{1}{c}{$N$} & \multicolumn{1}{c}{\textbf{GBT}} & \multicolumn{1}{c}{\textbf{CNN}} & \multicolumn{1}{c}{\textbf{CON}} 
\\ \hline 
\endfirsthead

\multicolumn{10}{c}%
{{\bfseries \tablename\ \thetable{} -- continued from previous page}} \\ \hline
\hline \multicolumn{1}{c}{\textbf{PDB ID}} & \multicolumn{1}{c}{$N$} & \multicolumn{1}{c}{\textbf{GBT}} & \multicolumn{1}{c}{\textbf{CNN}} & \multicolumn{1}{c|}{\textbf{CON}} &
\multicolumn{1}{c}{\textbf{PDB ID}} & \multicolumn{1}{c}{$N$} & \multicolumn{1}{c}{\textbf{GBT}} & \multicolumn{1}{c}{\textbf{CNN}} & \multicolumn{1}{c}{\textbf{CON}} 
\\ \hline 
\endhead

\hline \multicolumn{3}{|r|}{{Continued on next page}} \\ \hline
\endfoot

\hline \hline
\endlastfoot

1ABA&	87&	0.73&	0.71&	0.74&	2X5Y&	185&	0.76&	0.68&	0.76\\
1AHO&	66&	0.66&	0.66&	0.7&	2X9Z&	266&	0.49&	0.52&	0.52\\
1AIE&	31&	0.75&	0.7&	0.78&	2XHF&	310&	0.58&	0.57&	0.58\\
1AKG&	16&	0.27&	0.32&	0.29&	2Y0T&	111&	0.71&	0.71&	0.74\\
1ATG&	231&	0.55&	0.51&	0.55&	2Y72&	183&	0.65&	0.71&	0.69\\
1BGF&	124&	0.61&	0.58&	0.62&	2Y7L&	323&	0.66&	0.66&	0.68\\
1BX7&	51&	0.74&	0.74&	0.76&	2Y9F&	149&	0.74&	0.75&	0.76\\
1BYI&	238&	0.61&	0.5&	0.6&	2YLB&	418&	0.67&	0.66&	0.7\\
\end{longtable}}

\begin{table}
\centering
\caption{Pearson correlation coefficients of  least squares fitting C$_{\alpha}$ B factor prediction of small proteins using 11\AA\ cutoff. Two Bottleneck (B) and Wasserstein (W) metrics using various kernel choices are included.}
\label{tab:small}
\begin{tabular}{cc|ccc|ccc|ccc}\hline\hline
&  & \multicolumn{3}{c|}{B \& W}& \multicolumn{3}{c|}{B} 	& \multicolumn{3}{c}{W}\\ \hline 
PDB ID &$N$&	Exp	&	Lor	&	Both&Exp	&	Lor	&	Both	&	Exp	&	Lor	&	Both	\\ \hline
1AIE&31	&0.97&0.88&	0.99&	0.78	&	0.64	&	0.90	&	0.90&	0.77&0.96 \\
1AKG&16	&	0.82	&	0.66	&	1.00	&	0.60	&	0.53	&	0.72	&	0.53&0.56	&	0.87\\
1BX7&51	&	0.86	&	0.74	&	0.89	&	0.79	&	0.68	&	0.82	&	0.81&0.69	&	0.82\\
1ETL&12	&	1.00	&	1.00	&	1.00	&	0.68	&	0.87	&	1.00	&	0.95&0.98	&	1.00\\
1ETM&12	&	1.00	&	1.00	&	1.00	&	0.45	&	0.74	&	0.86	&	0.70&0.83	&	1.00\\
1ETN&12	&	1.00	&	1.00	&	1.00	&	0.96	&	0.92	&	0.99	&	0.70&0.92	&	1.00\\
\hline \hline

\end{tabular}
\end{table}

\begin{table}
\centering
\caption{Pearson correlation coefficients of  least squares fitting C$_{\alpha}$ B factor prediction of medium proteins using 11\AA\ cutoff. Two Bottleneck (B) and Wasserstein (W) metrics using various kernel choices are included.}
\label{tab:med}
\begin{tabular}{cc|ccc|ccc|ccc}\hline\hline
&  & \multicolumn{3}{c|}{B \& W}& \multicolumn{3}{c|}{B} 	& \multicolumn{3}{c}{W}\\ \hline 
PDB ID &$N$&	Exp	&	Lor	&	Both&Exp	&	Lor	&	Both	&	Exp	&	Lor	&	Both	\\ \hline
1ABA	&	87	&	0.67	&	0.67	&	0.76	&	0.54	&0.62&0.68&0.56	&	0.63&0.70\\
1CYO	&	88	&	0.71	&	0.69	&	0.78	&	0.66	&0.58	&	0.68	&	0.65	&	0.59	&	0.67\\
1FK5	&	93	&	0.53	&	0.59	&	0.71	&	0.49	&0.50	&	0.58	&	0.49	&	0.50	&	0.55\\
1GXU	&	89	&	0.75	&	0.78	&	0.82	&	0.72	&0.61	&	0.75	&	0.69	&	0.72	&	0.77\\
1I71	&	83	&	0.44	&	0.66	&	0.76	&	0.41	&0.46	&	0.56	&	0.38	&	0.58	&	0.59\\
1LR7	&	73	&	0.61	&	0.62	&	0.71	&	0.57	&0.55	&	0.63	&	0.46	&	0.56	&	0.58\\
\hline\hline
\end{tabular}
\end{table}

\begin{table}
\centering
\caption{Pearson correlation coefficients of  least squares fitting C$_{\alpha}$ B factor prediction of large proteins using 11\AA\ cutoff. Two Bottleneck (B) and Wasserstein (W) metrics using various kernel choices are included.}
\label{tab:lar}
\begin{tabular}{cc|ccc|ccc|ccc}\hline\hline
&  & \multicolumn{3}{c|}{B \& W}& \multicolumn{3}{c|}{B} 	& \multicolumn{3}{c}{W}\\ \hline 
PDB ID &$N$&	Exp	&	Lor	&	Both&Exp	&	Lor	&	Both	&	Exp	&	Lor	&	Both	\\ \hline
1AHO	&	66	&	0.75	&	0.78	&	0.88	&	0.72	&	0.73	&	0.79	&	0.53	&	0.65	&	0.75	\\
1ATG	&	231	&	0.50	&	0.50	&	0.61	&	0.45	&	0.47	&	0.53	&	0.38	&	0.48	&	0.51	\\
1BYI	&	238	&	0.50	&	0.51	&	0.58	&	0.41	&	0.46	&	0.49	&	0.44	&	0.48	&	0.54	\\
1CCR	&	109	&	0.65	&	0.66	&	0.71	&	0.53	&	0.56	&	0.65	&	0.43	&	0.58	&	0.63	\\
1E5K	&	188	&	0.67	&	0.68	&	0.74	&	0.66	&	0.67	&	0.68	&	0.63	&	0.67	&	0.69	\\
1EW4	&	106	&	0.58	&	0.60	&	0.73	&	0.52	&	0.51	&	0.55	&	0.55	&	0.55	&	0.62	\\
\end{tabular}
\end{table}

\begin{longtable}{*{11}{c}}
\caption{Pearson correlation coefficients of least squares fitting C$_{\alpha}$ B factor prediction of all proteins using 11\AA\ cutoff. Two Bottleneck (B) and Wasserstein (W) metrics using various kernel choices are included. } \label{tab:all}\\
\hline\hline \multicolumn{1}{c}{}& \multicolumn{3}{c}{B \& W}& \multicolumn{3}{c}{B}& \multicolumn{3}{c}{W}\\ \hline 
PDB ID   & $N$&	Exp	&	Lor	&	Both	&	Exp	&	Lor	&	Both	&	Exp	&	Lor	&	Both\\ \hline
\endfirsthead
\multicolumn{11}{c}%
{{\bfseries \tablename\ \thetable{} -- continued from previous page}} \\ \hline
\hline \multicolumn{1}{c}{}& \multicolumn{3}{c}{B \& W}& \multicolumn{3}{c}{B}& \multicolumn{3}{c}{W}\\ \hline 
PDB ID   & $N$&	Exp	&	Lor	&	Both	&	Exp	&	Lor	&	Both	&	Exp	&	Lor	&	Both\\ \hline
\endhead
1ABA	&	87	&	0.67	&	0.67	&	0.76	&	0.54	&	0.62	&	0.68	&	0.56	&	0.63	&	0.70	\\
1AHO	&	66	&	0.75	&	0.78	&	0.88	&	0.72	&	0.73	&	0.79	&	0.53	&	0.65	&	0.75	\\
1AIE	&	31	&	0.97	&	0.88	&	0.99	&	0.78	&	0.64	&	0.90	&	0.90	&	0.77	&	0.96	\\
1AKG	&	16	&	0.82	&	0.66	&	1.00	&	0.60	&	0.53	&	0.72	&	0.53	&	0.56	&	0.87	\\
1ATG	&	231	&	0.50	&	0.50	&	0.61	&	0.45	&	0.47	&	0.53	&	0.38	&	0.48	&	0.51	\\
1BGF	&	124	&	0.75	&	0.70	&	0.82	&	0.64	&	0.54	&	0.75	&	0.68	&	0.61	&	0.75	\\
1BX7	&	51	&	0.86	&	0.74	&	0.89	&	0.79	&	0.68	&	0.82	&	0.81	&	0.69	&	0.82	\\
1BYI	&	238	&	0.50	&	0.51	&	0.58	&	0.41	&	0.46	&	0.49	&	0.44	&	0.48	&	0.54	\\
\end{longtable}

%

\vspace{1cm}
\clearpage


\newpage



\end{document}